\documentclass[epj]{svjour}
\usepackage{tikz}
\usetikzlibrary{patterns}
\usepackage{axodraw}
\usepackage{cite}
\usepackage{epsfig,multicol}
\usepackage{lineno}
\usepackage{amsmath,amssymb,mathrsfs}

\newcommand\lsim{\mathrel{\raise.3ex\hbox{$<$\kern-.75em\lower1ex\hbox{$\sim$}}}}
\newcommand\gsim{\mathrel{\raise.3ex\hbox{$>$\kern-.75em\lower1ex\hbox{$\sim$}}}}

\newenvironment{Eqnarray}%
     {\arraycolsep 0.14em\begin{eqnarray}}{\end{eqnarray}}
\newcommand{\beqa}{\begin{Eqnarray}}
\newcommand{\eeqa}{\end{Eqnarray}}
\newcommand{\beq}{\begin{equation}}
\newcommand{\eeq}{\end{equation}}
\newcommand{\tr}{\mathrm T \mathrm r}  
\begin{document}
\title{
Probing Compositeness with 
Higgs Boson Decays at the LHC}
\author{Maria Hoffmann\inst{1}, Anna Kaminska\inst{2}, Rosy Nicolaidou\inst{1} and Stathes Paganis\inst{3}\footnote{Part of the work performed at the Department of Physics and Astronomy, University of Sheffield, Sheffield, United Kingdom}} 
 
\institute{DSM/IRFU (Institut de Recherches sur les Lois Fondamentales de l'Univers), CEA Saclay (Commissariat a l'Energie Atomique), Gif-sur-Yvette, France
  \and Deutsches Electronen-Synchrotron DESY, Hamburg, Germany
  \and Department of Physics, National Taiwan University, Taipei, Taiwan
}
%


\abstract{
A method is proposed to directly probe the Higgs boson compositeness 
using the unique characteristics of a boosted Higgs boson produced in
association with a weak gauge boson ($W^{\pm},Z$).
The discovery potential for the upcoming LHC running 
is presented, showing that compositeness scales up to 3~TeV can be probed at the LHC 
with an integrated luminosity of $\mathcal{L}=3000$~fb$^{-1}$ collected at $\sqrt{s}=13$ TeV.
} 
\titlerunning {Probing Compositeness with Higgs Boson Decays at the LHC}
\authorrunning{M.Hoffmann, A.Kaminska, R.Nicolaidou, S.Paganis} 
\maketitle

\section{Introduction}
\label{intro}
After the discovery of a particle consistent with the Standard Model
(SM) Higgs boson \cite{BEH1,BEH2,BEH3}
by the 
ATLAS and CMS Collaborations \cite{CMS,ATLAS}, intense research for the understanding
of the details of the Higgs mechanism has commenced. Experimental data does not rule out a possible composite nature of 
the Higgs boson. 
In composite Higgs models, the Higgs boson is a (pseudo) Goldstone boson emerging as a result of spontaneously broken global symmetry of a new, strong dynamics. Like in QCD, there is a new fermion sector 
causing spin-1 $\rho $-like bound states with masses  
at the compositeness energy scale.

The $\rho$-like bound states ($\rho^{0}$, $\rho^{+},\rho^{-}$) couple to SM
particles and can hence be directly probed at the LHC through searches 
for single lepton events ($\rho^{\pm} \rightarrow \ell^{\pm} \nu$) 
and searches for resonances decaying to two ($\rho^{0} \rightarrow \ell^{+} \ell^{-}$) and three charged leptons ($\rho^{\pm} \rightarrow W^{\pm} Z \rightarrow \ell^{\pm} \nu \ell^{+} \ell^{-}$).

Currently, the dilepton searches set the best limits on
compositeness \cite{dileptonCMS, dileptonATLAS}.
However, for a large part of the compositeness model phase space, 
the branching ratio (BR) of $\rho^0\rightarrow \ell^{+} \ell^{-}$ falls very fast with 
the $\rho$ mass. Meanwhile, the BR of $\rho\rightarrow
VH$, where V$=$W$^{\pm}$,Z, reaches a maximum. 
 An observation of VH events 
in excess of what is expected  by the SM
and with an invariant mass at the TeV scale would be strong
evidence for compositeness.\\
\indent The decay products 
originating from heavy $\rho$ decays are characterised by a very high transverse momentum ($p_{\perp}$). 
By exploiting this feature we propose a method to explore the presence of compositeness via the search for
Higgs bosons with high $p_{\perp}$. 
This search is complementary to current searches using 
the dilepton invariant mass, and it may be more powerful than the latter in the parts of the compositeness 
model phase space where the BR$\left( \rho^0 \rightarrow \ell^{+} \ell^{-} \right) \lesssim 0.5\% $.\\
\indent In this work we search for Higgs boson decays in the two channels providing the experimentally cleanest signatures for reconstructing the 
Higgs; the ``golden'' decay to four leptons,
$H\rightarrow ZZ^{*} \rightarrow 4 \ell$, where $\ell = e, \mu$, and the 
decay to two photons, $H\rightarrow \gamma\gamma$. 
The addition of the $H\rightarrow b\bar{b}$ mode is also discussed. \\
\indent Since a Higgs originating from heavy $\rho$ decays will carry a large $p_{\perp}$, all non-SM backgrounds for 
$H \rightarrow ZZ^{*} \rightarrow 4 \ell$ and $H\rightarrow \gamma\gamma$ decays
are expected to be small and SM Higgs backgrounds will thus dominate. 
In the $H\rightarrow b\bar{b}$ channel the non-Higgs background is still significant. 
The SM backgrounds can
be strongly suppressed with a high Higgs $p_{\perp}$ requirement, which is
the approach employed in this work. With the proposed method we are able to 
set a direct limit on the existence of compositeness and its energy scale. 


\section{Spin-1 Resonances as a Signal of Composite Higgs} 
\label{theory}
In this section we examine theories where the Higgs boson is
a composite pseudo-Nambu-Goldstone boson (PGB) \cite{Contino}.
In these theories, a new strongly interacting sector with global symmetry $\mathcal{G}$
is present at the $\gtrsim 1$~TeV compositeness scale ($f$). A composite 
Higgs boson emerges, much like the pion of QCD, as the PGB of a global 
symmetry breaking $\mathcal{G} \rightarrow \mathcal{H}$ of that sector. 
The explicit symmetry breaking is induced by interactions of 
the SM gauge bosons and fermions
with the strong sector.

\begin{figure}[htb]
\resizebox{0.45\textwidth}{!}{
\includegraphics{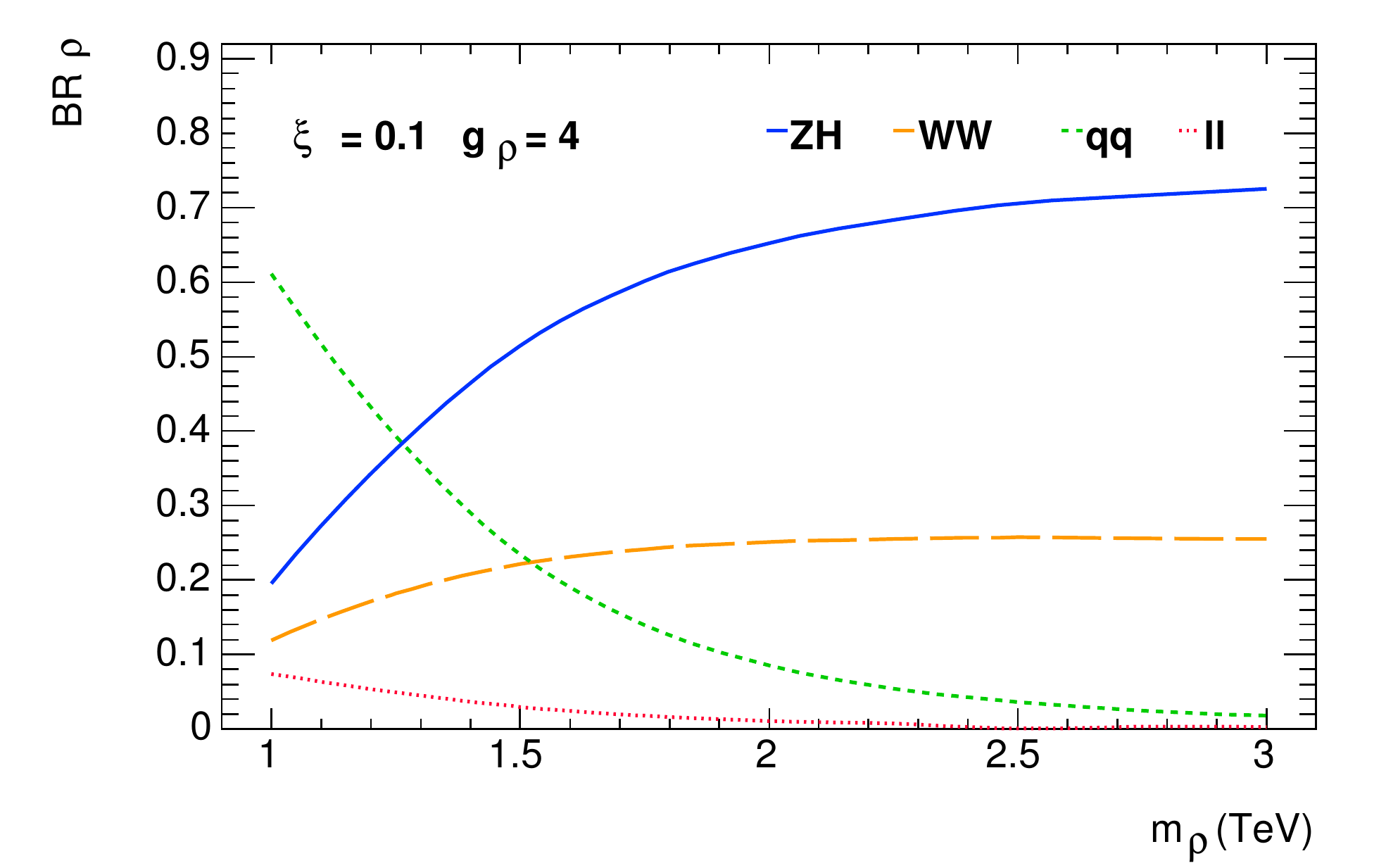}
}
\resizebox{0.45\textwidth}{!}{
\includegraphics{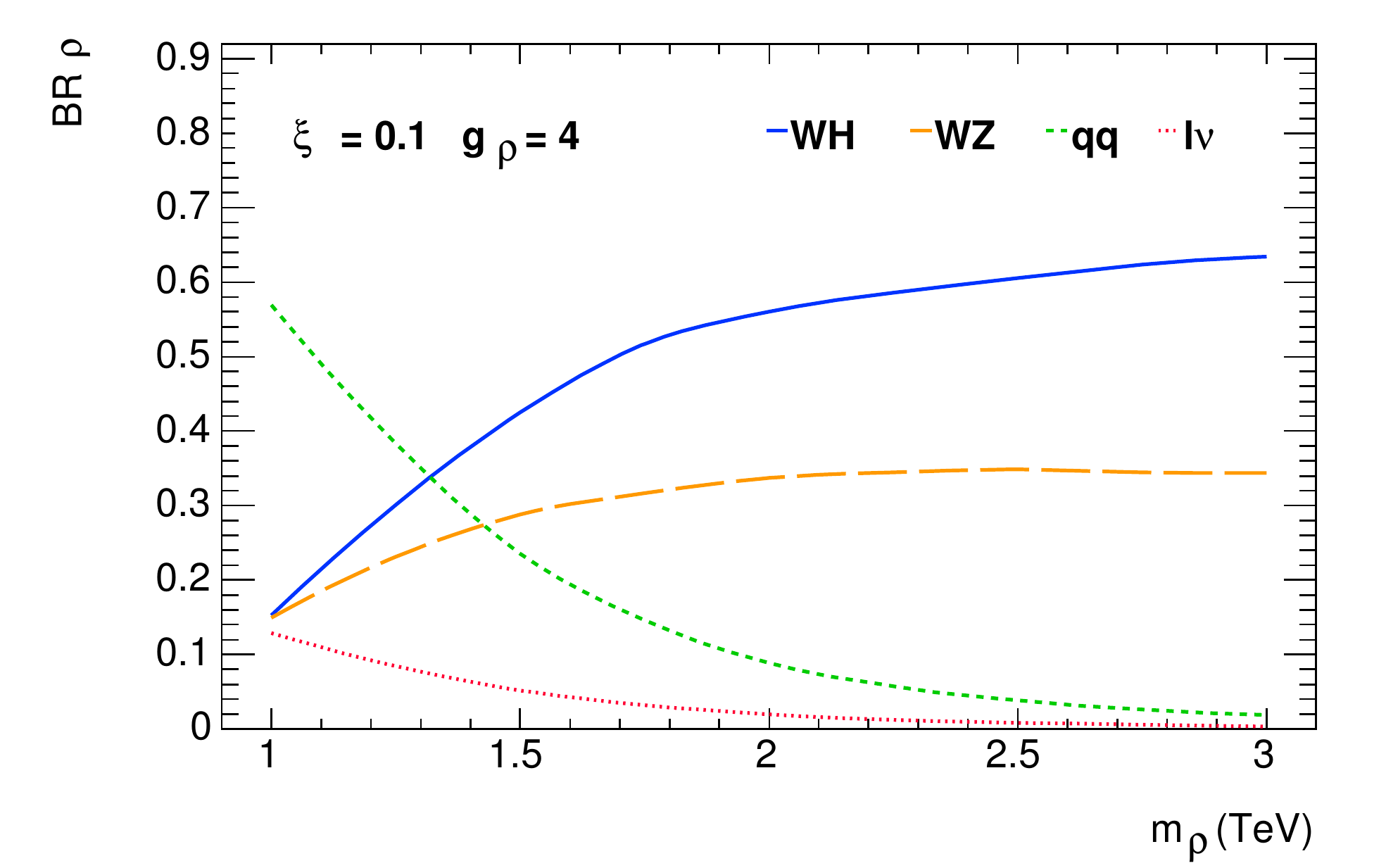}
}
\caption{Branching ratios of $\rho^{0}$ (top) and $\rho^{\pm}$ (bottom) decay modes for the benchmark
  parameters $\xi = 0.1$ and
  $g_{\rho}=4$. The dilepton BR drops quickly to 
zero as a function of $m_{\rho}$ while the BR of $\rho \rightarrow VH$   
tends to a maximum of about 70\%.
 }
\label{rhoBR}

\end{figure}

The simplest example of such a strong sector 
is $SO(5) \rightarrow SO(4)$ \cite{Contino}, where 
$SO(4) \sim SU(2)_L \times SU(2)_R$, and the 
$10-6=4$
pseudo-Goldstones form a complex scalar field
$SU(2)_L$ doublet that plays the role of the Higgs. 
The $SO(4) \sim SU(2)_L \times SU(2)_R $
global symmetry is gauged by the electroweak symmetry $\mathcal{G}_{SM}=SU(2)_L \times U(1)_Y$ of the
SM, which is external to the new strong sector. This means that the 
SM gauge bosons are external and couple to the strong sector.
Interactions of the SM gauge bosons and fermions with the strong sector
are responsible for the explicit breaking of the global symmetry
$\mathcal{G}$. 
In this picture, loops of SM fermions and gauge bosons generate a Higgs potential 
which eventually breaks electroweak symmetry at scale $v$. 
This dynamically generated electroweak scale $v$ may be lower 
than the strong sector (compositeness) breaking scale $f$.
The ratio between the two scales $\xi=\left(v/f\right)^2$ is determined by the orientation 
of the electroweak group $\mathcal{G}_{SM}$ with respect to the unbroken 
strong sector group $\mathcal{H}=SU(2)_L \times SU(2)_R$ in
the true vacuum. If these two groups are misaligned, the 
electroweak symmetry is broken.
Three composite pseudo-Goldstones become the longitudinal degrees of freedom 
of the weak gauge bosons and the fourth PGB defined along the
misalignment angle $\theta$ is the light Higgs boson. \\
\begin{figure}[t!]
\begin{center}
\resizebox{0.3\textwidth}{!}{
\includegraphics{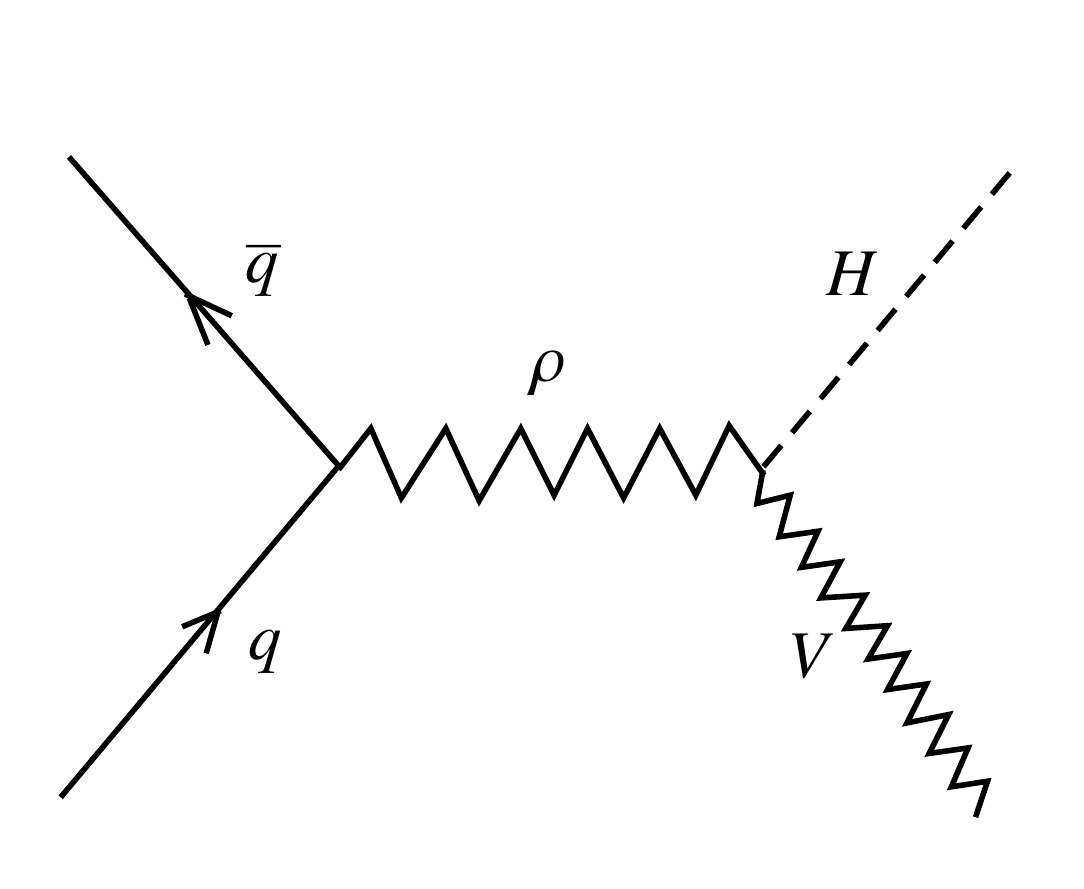}
}
\caption{Diagram depicting the process $pp \rightarrow \rho \rightarrow VH$.}
\label{FMdiag}
\end{center}
\vspace{-5mm}
\end{figure}
\indent In such composite Higgs models, vector meson $\rho$-like resonances
appear and mix with the gauge bosons building new spin-1 eigenstates.
Fermion resonances from the strong sector mix with SM fermions allowing them to interact directly with
$\rho$ resonances. \\
\indent In this paper we consider the simplest case of the 
Minimal Composite Higgs Model (MCHM) \cite{Contino}. 
The phenomenology of $\rho$ resonances in this model, transforming in the adjoint representation of $SU(2)_L$, is
representative of the entire family of composite Higgs models. It is expected that $SU(2)_L\times SU(2)_R\subset \mathcal{H}$ for any $\mathcal{G}\rightarrow\mathcal{H}$ composite Higgs model, in order to have custodial symmetry embedded in the construction. Since vector meson resonances appear in representations of the unbroken global group $\mathcal{H}$, $\rho$ resonances transforming as $(3,1)$ of $SU(2)_L\times SU(2)_R$ are a generic prediction of theories with a composite Higgs boson. Such resonances are expected to have substantial interactions with SM particles due to their natural mixing with $W_{\mu}$ fields. General properties of $SU(2)_L$ $\rho$ mesons are well described in the framework of MCHM. \\
\indent In the following analysis we consider $\rho$ resonances transforming as a triplet of $SU(2)_L$, generically expected in composite Higgs models as discussed above. For the effective description of spin-1 resonances we follow the CCWZ approach \cite{Coleman:1969sm, Callan:1969sn}, just as is done in \cite{Contino:2011np} and presented in the Appendix. 
This approach is fully equivalent to the Hidden Local Symmetries formalism \cite{Bando:1987br}. Hence it is compatible with any generic composite Higgs construction used in the literature on the level of effective Lagrangian description. Moreover, it has a direct connection with deconstruction of extra dimensions, as discussed in \cite{Panico:2011pw}. \\
\indent When mixing effects in the spin-1/2 sector are neglected,
compositeness models with a single vector meson can be described by just three parameters: the $\rho$
mass, $m_{\rho}$, the $\rho$ self-coupling $g_{\rho}$ and the parameter $\xi$. 
Based on naive dimensional analysis (NDA), the mass scale of $\rho$ resonances is expected to be $m_{\rho}\sim g_{\rho} f$, where $1<g_{\rho}<4\pi$.
The $\rho$ production cross section is dominated by the Drell-Yan 
process. Neutral $\rho$ decays to dileptons are particularly appealing 
to experimental searches for compositeness at the LHC. In particular 
for small $m_{\rho} (<2$~TeV) and small values of parameter 
$\xi (\lesssim 0.1$) the BR of $\rho^0\rightarrow \ell^+\ell^-$ 
is large. However, observation of an excess in this channel would not by 
itself be sufficient to claim the observation of a composite $\rho$. Observation 
of the rest of the modes shown in Fig.~\ref{rhoBR} would be required.\\
\indent Electroweak precision observables (see for example the discussion in \cite{Contino:2013gna}) and LHC collider
data are sensitive to compositeness and can set limits in parts of the parameter space. 
Significant enhancement in the $H\rightarrow Z\gamma$ yield can also be induced 
by compositeness effects \cite{ZgComp}. 
ATLAS and CMS have set limits on the $\xi$ parameter using Higgs couplings
$\xi<0.22$ at 95\% CL, restricting the compositeness scale to 
$f>0.5$~TeV
\cite{ATLAS-CONF-2014-10, mor13}. It is safe to say that $\xi=0.1$ at this point is still consistent with both electroweak precision constraints and LHC Higgs data, hence it will be used as a benchmark value in this paper. The compositeness scale can be also probed by direct searches for lightest vector resonances, though there is no strict relation between $m_{\rho}$ and $f$. However, as mentioned before, by NDA we can expect $m_{\rho}\sim g_{\rho}f$. Using this assumption, the LHC narrow mass dilepton searches place limits in the region of $f\lesssim 1.5$~TeV$/g_{\rho}$ \cite{dileptonCMS, dileptonATLAS}. These searches are most efficient for exploring the light $\rho$ resonance ($m_{\rho}\lesssim 2$~TeV) parameter space with substantial branching ratios into lepton pairs. 
However, for a big part of the parameter space the dilepton BR drops quickly to 
zero as a function of $m_{\rho}$ while the BR of $\rho \rightarrow VH$ 
tends to a maximum of about 70\%. The $\rho$ BRs as 
a function of the $m_\rho$ are shown in Figure \ref{rhoBR} for the benchmark 
parameter values of $\xi=0.1$ and $g_{\rho}=4$. 
Naively one would expect the BRs into VV and into VH to be equal. However, the fact that the cosine of the Weinberg angle $\cos\theta_W$ is not equal 1 leads to substantial differences in the $\rho$ couplings to $W$ and $Z$ bosons in some parts of the parameter space (especially for $g_{\rho}\lesssim 6$).
For $m_{\rho}>2$~TeV, the BR$\left(\rho^0\rightarrow \ell^+\ell^-\right)$ drops to 
values below 1\%.
The branching ratios of $\rho$ mesons do not depend strongly on the choice of $g_{\rho}$, but the production cross section behaves roughly as $1/g_{\rho}^2$, hence exclusion limits for $\rho$ are $g_{\rho}$ dependent. In general it is expected that $g_{\rho}$ is substantially larger than the weak couplings $g,\; g'$. The $\rho$ meson of QCD is described by $g_{\rho}\sim 6$. In this paper we consider $g_{\rho}=4$ as a benchmark value, which is sufficiently above the weak couplings scale but still allows for significant $\rho$ production at the LHC.\\
\indent The large expected VH branching ratio of the $\rho$ meson and the fact that both charged and neutral
$\rho$'s are involved offers a new possibility in experimental searches. We propose to
search for boosted Higgs bosons produced by $\rho^0,\rho^+,\rho^-$ decays 
in association with gauge bosons. For
BR$\left(\rho^0\rightarrow\ell^+\ell^-\right) 
\gtrsim 0.5$\% the proposed search is complementary to the
dilepton search and can add information on the origin of a potential
excess seen in the dilepton mass spectrum. For smaller 
BR$\left(\rho^0\rightarrow\ell^+\ell^-\right)$, the VH search becomes 
the most powerful in exploring the compositeness parameter space. In addition, a salient feature of the VH decay is that the Higgs invariant mass may be used to suppress the background.
%
%
\section{Monte Carlo Samples} 
The results presented in this paper are based on Monte Carlo (MC) samples generated with MadGraph5~\cite{madgraph5} and parton showered with PYTHIA~\cite{Pythia8:2008}. To simulate the response of an LHC-like experiment, realistic resolution and reconstruction efficiencies for electrons, muons, photons and jets were applied with the Delphes framework \cite{delphes}. \\
\indent The signal samples include the processes $pp\rightarrow \rho \rightarrow VH$, 
where the definitions from Section \ref{intro} apply. 
These samples were generated  at $\sqrt{s} = 13$ TeV with the benchmark parameter values $\xi = 0.1$ and $g_{\rho}=4$.
Typical values for the $\rho$ production cross section as a function of the $\rho$ mass 
computed with MadGraph5 are presented in Table \ref{xsections13}. Numerical values of the BR for the processes $\rho^0\rightarrow ZH$,
$\rho^0\rightarrow \ell^+\ell^-$, $\rho^{\pm}\rightarrow W^{\pm}H$ and $\rho^{\pm}\rightarrow \ell^{\pm}\nu$  are shown in Table~\ref{br13}.

The Higgs boson in the signal samples decays via either of the two channels $H \rightarrow \gamma \gamma $ or $H \rightarrow ZZ^{*} \rightarrow 4 \ell$, where $\ell =e,\mu$. The vector boson produced in association with the Higgs boson is constrained to decay hadronically, i.e. $V \rightarrow jj$. \\
\indent As already mentioned, the main source of background in this search is SM Higgs production. In this study we consider the three main Higgs production mechanisms at the LHC; gluon-gluon fusion (ggF), vector boson fusion (VBF) and associated production (WH/ZH)\footnote{The cross-section of a  Higgs boson produced in association with a pair of top or b quarks ($t\bar{t}H$ or $b\bar{b}H$) is $\approx$ 3 times lower than the WH/ZH production at $\sqrt{s}=13$ TeV and therefore not considered in this study.}. All samples were generated with a Higgs mass of $m_{H} = 125$ GeV, and scaled with the relevant cross sections and branching ratios reported by the LHC Higgs cross section working group and the Particle Data Group \cite{higgsXS,PDG}. \\
\indent In the following we further assess the improvement of the search when adding the high BR channel $H\rightarrow b\bar{b}$. This is done by extrapolating the results obtained for the $H \rightarrow \gamma \gamma$ channel using results from a recent combination reported by the LHC experiments in~\cite{ttH, Hbb}.
%
%
\begin{table}
\small
  \begin{center}    
    \begin{tabular}{ccccc}\hline
      m$_{\rho}$ [TeV] &$\sigma_{\rho^0}$ [fb] & $\sigma_{\rho^+}$
      [fb] & $\sigma_{\rho^-}$ [fb] & Total [fb]\\\hline
   1.50 & 59.11 & 92.27& 32.94& 184.3 \\
   1.75 & 27.92 & 45.25& 14.91& 88.08 \\      
   2.00 & 13.94 & 23.35& 7.172& 44.46 \\      
   2.25 & 7.245 & 12.48& 3.606& 23.33 \\      
  2.50 & 3.873 & 6.835& 1.874& 12.58 \\      
   2.75 & 2.121 & 3.807& 0.992&  6.920 \\     
  3.00 & 1.118 & 2.144& 0.543& 3.805 \\   
    \hline
    \end{tabular}
    \caption{ Total and individual cross sections for the process $pp \rightarrow \rho$ as function of $m_{\rho}$. The cross sections were computed at $\sqrt{s}=13$~TeV with the parameter values $\xi=\left(v/\Lambda_c\right)^2=0.1$, $g_{\rho}=4$. The statistical uncertainty is less than $10^{-7}$ fb, and therefore not quoted in this table.} 
    \label{xsections13}
  \end{center}
\end{table}
\begin{table}
\small
  \begin{center}    
  \resizebox{0.45\textwidth}{!}{
    \begin{tabular}{ccccc}\hline
      m$_{\rho}$ [TeV] &$\rho^0\rightarrow ZH$  &
      $\rho^0\rightarrow \ell^+\ell^-$ & $\rho^{\pm} \rightarrow W^{\pm}H$ &$\rho^{\pm}\rightarrow \ell^\pm\nu$\\
      \hline
      1.50 & 0.515 & 0.0297   & 0.416  & 0.0716   \\
      1.75 & 0.603 & 0.0179   & 0.499  & 0.0442   \\
      2.00 & 0.653 & 0.0110   & 0.551  & 0.0277   \\
      2.25 & 0.683 & 0.00703 & 0.584  & 0.0180   \\
      2.50 & 0.704 & 0.00469 & 0.604  & 0.0120   \\
      2.75 & 0.714 & 0.00316 &  0.617 & 0.00833 \\
      3.00 & 0.725 & 0.00229 & 0.627  & 0.00593 \\
      \hline
    \end{tabular}
    }
    \caption{Branching ratios for the decays $\rho^0\rightarrow ZH \rightarrow \ell^+\ell^-H$, $\rho^{\pm} \rightarrow W^{\pm}H$ and $\rho^{\pm} \rightarrow \ell^{\pm}\nu$ as function of $m_{\rho}$. The branching ratios were computed with the parameter values $\xi=\left(v/\Lambda_c\right)^2=0.1$, $g_{\rho}=4$.} 
    \label{br13}
  \end{center}
\end{table}
%
%
%
\section{Analysis Strategy}\label{sec:analysis} 
\begin{figure}[htb]
\resizebox{0.45\textwidth}{!}{
\includegraphics{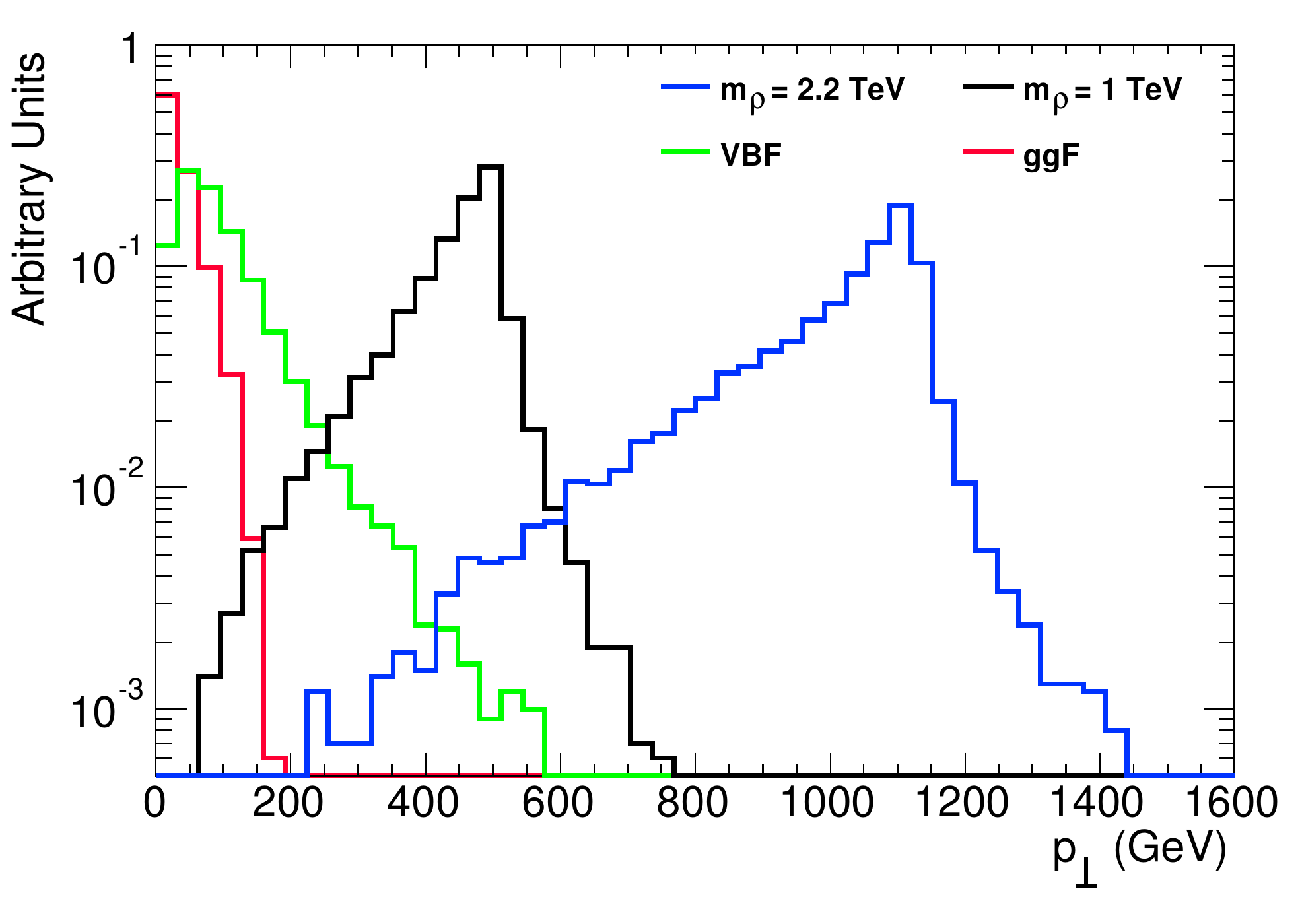}
}
\caption{
The distribution of transverse momentum $p_{\perp}$ of SM Higgs bosons and of 
Higgs bosons originating from the $\rho\rightarrow VH$ decay.}
\label{higgspt}
%

\vspace{5mm}
\resizebox{0.45\textwidth}{!}{%
\includegraphics{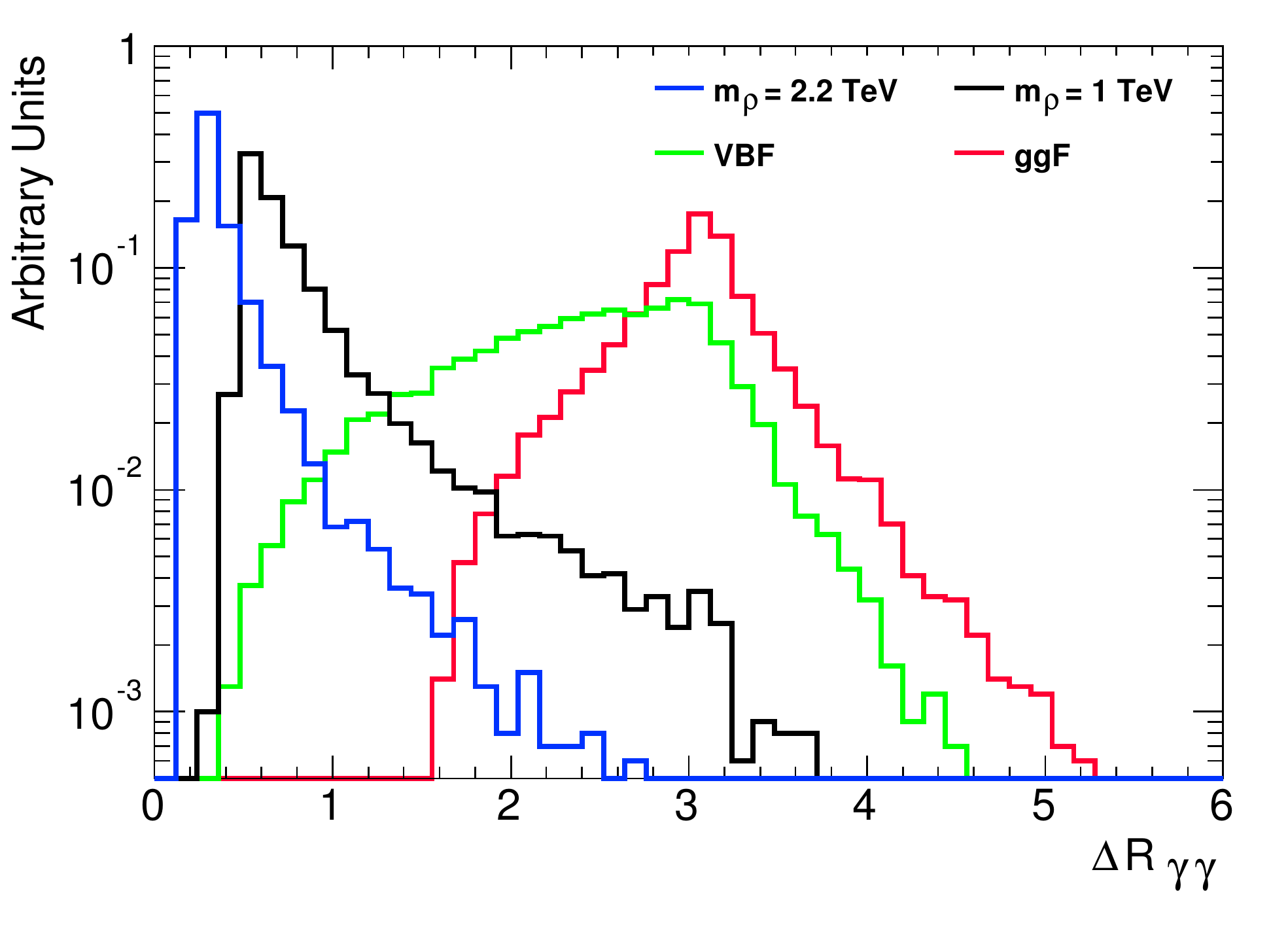}
}
\caption{The opening angle $\Delta R = \sqrt{ \Delta \eta^{2}  + \Delta \phi^{2}  }$ between two photons
from SM Higgs boson decays and from  $\rho \rightarrow VH \rightarrow V \gamma \gamma$ decays.}
\label{ggDR}
\end{figure}

The analysis method presented in this paper aims towards optimising the discovery potential of a composite Higgs during the upcoming LHC runs. A discovery could come in the form of a direct observation of Higgs boson events with anomalously high $p_{\perp}$ or $\rho$ decays to VH along with an excess of events with dilepton invariant mass at the TeV scale. \\
\indent 
The distribution of the transverse momentum of SM Higgs bosons 
and of Higgs bosons produced from $\rho$ decays is shown 
in Figure \ref{higgspt}. Given the noticeable difference in the shape of the distributions, 
the transverse momentum can be used as a discriminating variable to
suppress the SM background. 
%
The large transverse boost of these Higgs events causes 
the opening angle $\Delta R = \sqrt{ \Delta \eta^{2}  + \Delta \phi^{2}  }$
between the decay products ($ZZ$ and $\gamma\gamma$) 
to be significantly smaller than that from a SM Higgs, which is
 shown in Figure \ref{ggDR}. The characteristic $\Delta R$ is not exploited in the present analysis strategy, but may be used in future searches
to increase the sensitivity.

\subsection{Event Selection}\label{eventselection}
A simple event selection inspired by the analysis strategies of ATLAS and CMS is implemented. 
The $H\rightarrow\gamma\gamma$ signal is selected by requiring two hard photons 
with $E_{T}$ of the leading (subleading) photon being $\geq$ 40 (30) GeV.  Events from the $H \rightarrow ZZ^{*} \rightarrow 4 \ell$ decay are selected by requiring two pairs of oppositely charged, same-flavour leptons. The three leptons in the quadruplet with the largest transverse momentum must, in descending order, satisfy $p_{\perp}\geq$ 20, 15, 10 GeV.  Muons, electrons and photons must respectively satisfy $|\eta|<2.7$, 2.47, 2.37. Common for the channels is that the invariant mass of the Higgs final states\footnote{i.e. the $4\ell$ and $\gamma \gamma$ system} 
must be in the range 100 - 150 GeV. To suppress the SM backgrounds, an additional requirement on the transverse momentum
 of the Higgs system of $ p_{\perp}\geq$ 550 GeV is applied. \\ 
\indent 
Since the applied event selection is simplified with respect to that of the LHC experiments, all samples are multiplied with analysis efficiencies representative of those presented by ATLAS and CMS. These values were obtained at $\sqrt{s} = 8 $ TeV \cite{H4l,CMS-PAS-FTR-13-003}, and have been scaled to $\sqrt{s}=13$ TeV by assuming a slight increase in efficiency. A factor of 0.4 is applied to the signal samples in both Higgs decay channels. The factor applied to the $H \rightarrow ZZ^{*} \rightarrow 4 \ell$($H \rightarrow \gamma \gamma$) background sample is 0.3(0.4) for ggF, 0.2(0.5) for VBF, and 0.5(0.5) for WH/ZH. 
Furthermore, the signal is multiplied by a scale factor accounting for the increased detector fiducial acceptance, 
which is a consequence of the boost that causes the Higgs to be emitted at lower $\eta$.
The scale factor applied to  the $H \rightarrow \gamma \gamma$ channel is 1.2 while 1.3 is applied to $H \rightarrow ZZ^{*} \rightarrow 4 \ell$. \\
\indent The contribution from the non-resonant QCD background in the $\gamma \gamma+jj$ final state is estimated by extrapolating the obtained number of background events reported in \cite{Aad:2014eha} with the expected increase in production cross section \cite{ATL-PHYS-PUB-2013-014}, multiplied with the selection efficiency of the $p_{\perp}$ requirement. This efficiency is estimated with a  $pp \rightarrow \gamma \gamma+jj$ sample generated with MadGraph and found to be on the order of $10^{-4}$. With an integrated luminosity of 3000 fb$^{-1}$ the number of expected events from this process is more than one order of magnitude smaller than what is expected from the dominant SM Higgs background. This particular background is therefore considered safe to ignore. Similarly, the contribution from the $ZZ$ continuum ($pp \rightarrow ZZ$) is assumed to be negligible after the rather tight $p_{\perp}$ requirement. \\
\indent The number of signal and background events per fb$^{-1}$ remaining after the full selection are presented in Table \ref{yield1fbbkg} and \ref{yield1fbsig}. As seen, the
contribution from ggF production is completely suppressed. 
%
%
%
\begin{table}[h]
\small
\begin{center}
\begin{tabular}{ccc}
\hline
 SM production & $H \rightarrow ZZ^{*} \rightarrow 4 \ell$ & $H \rightarrow \gamma \gamma$  \\
 \hline
ggF  & 0 & 0 \\
VBF & $1.1 \cdot 10^{-4}$   & $5.2 \cdot 10^{-3}$ \\
ZH   & $4.0 \cdot 10^{-5}$   & $5.6 \cdot 10^{-4}$ \\
WH  & $6.3 \cdot 10^{-5}$   &  $8.8 \cdot 10^{-4}$\\
\hline
\end{tabular}
\caption{Number of background events remaining after the full selection in Section \ref{eventselection} with an integrated luminosity of $ \mathcal{L}=1$ fb$^{-1}$ at $\sqrt{s}=13$ TeV.} \vspace{5mm}\label{yield1fbbkg}
\begin{tabular}{ccc}
\hline
 m$_{\rho}$ [TeV] & $H \rightarrow ZZ^{*} \rightarrow 4 \ell$ & $H \rightarrow \gamma \gamma$  \\
 \hline
 1.50 & $3.8 \cdot 10^{-3}$   & $6.0 \cdot 10^{-2}$ \\
1.75 & $2.6 \cdot 10^{-3}$   & $3.9 \cdot 10^{-2}$ \\
2.00 & $1.5 \cdot 10^{-3}$   & $2.3 \cdot 10^{-2}$ \\
2.25 & $8.9 \cdot 10^{-4}$ & $1.5 \cdot 10^{-2}$\\
2.50 & $5.1 \cdot 10^{-4}$ & $8.3 \cdot 10^{-3}$ \\
2.75 & $2.9 \cdot 10^{-4}$ & $4.7 \cdot 10^{-3}$ \\
3.00 & $1.6 \cdot 10^{-4}$ & $2.6 \cdot 10^{-3}$\\
\hline
\end{tabular}
\caption{Number of signal events remaining after the full selection in Section \ref{eventselection} with an integrated luminosity of $ \mathcal{L}=1$ fb$^{-1}$ at $\sqrt{s}=13$ TeV.} \vspace{5mm}\label{yield1fb}
\label{yield1fbsig}
\end{center}
\vspace{-15mm}
\end{table}
%
%
\section{Results: Discovery Potential at the LHC}
\begin{figure}[htb]
\resizebox{0.49\textwidth}{!}{%
\includegraphics{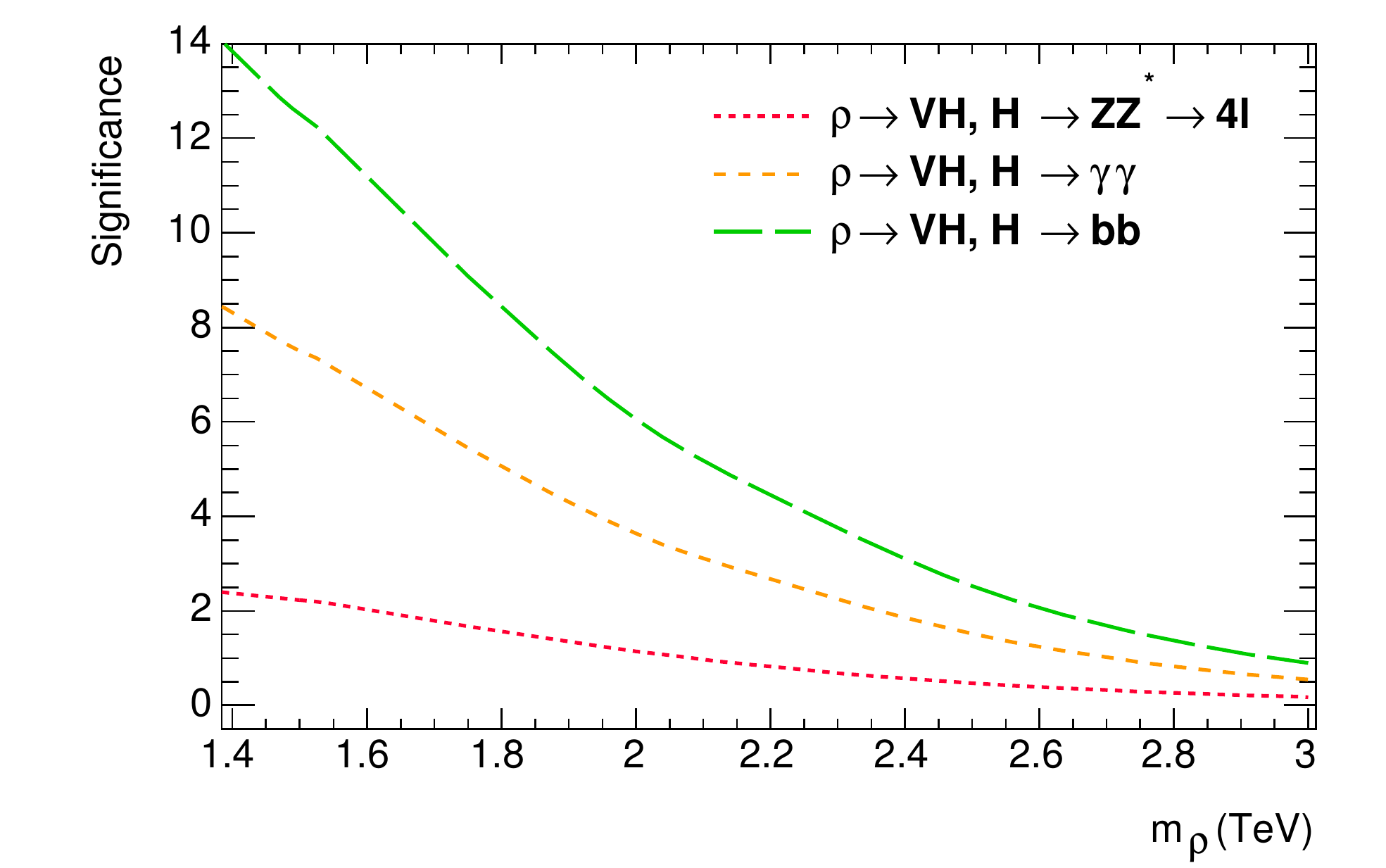}
}
\caption{
Expected significance as function of $m_{\rho}$ obtainable with a single LHC experiment using the $H\rightarrow \gamma\gamma$, $H\rightarrow ZZ^{*} \rightarrow 4\ell$ and $H \rightarrow b\bar{b}$ decay channels. The latter was obtained with an optimistic extrapolation of the results obtained with the $H\rightarrow \gamma\gamma$ channel with the results reported in \cite{ttH, Hbb}. The significance was computed with an integrated luminosity of $\mathcal{L}=300$ fb$^{-1}$ collected at $\sqrt{s}=13$ TeV with the parameter values $\xi=\left(v/\Lambda_c\right)^2=0.1$, $g_{\rho}=4$.}

\label{signif300}
\vspace{5mm}
\resizebox{0.49\textwidth}{!}{%
\includegraphics{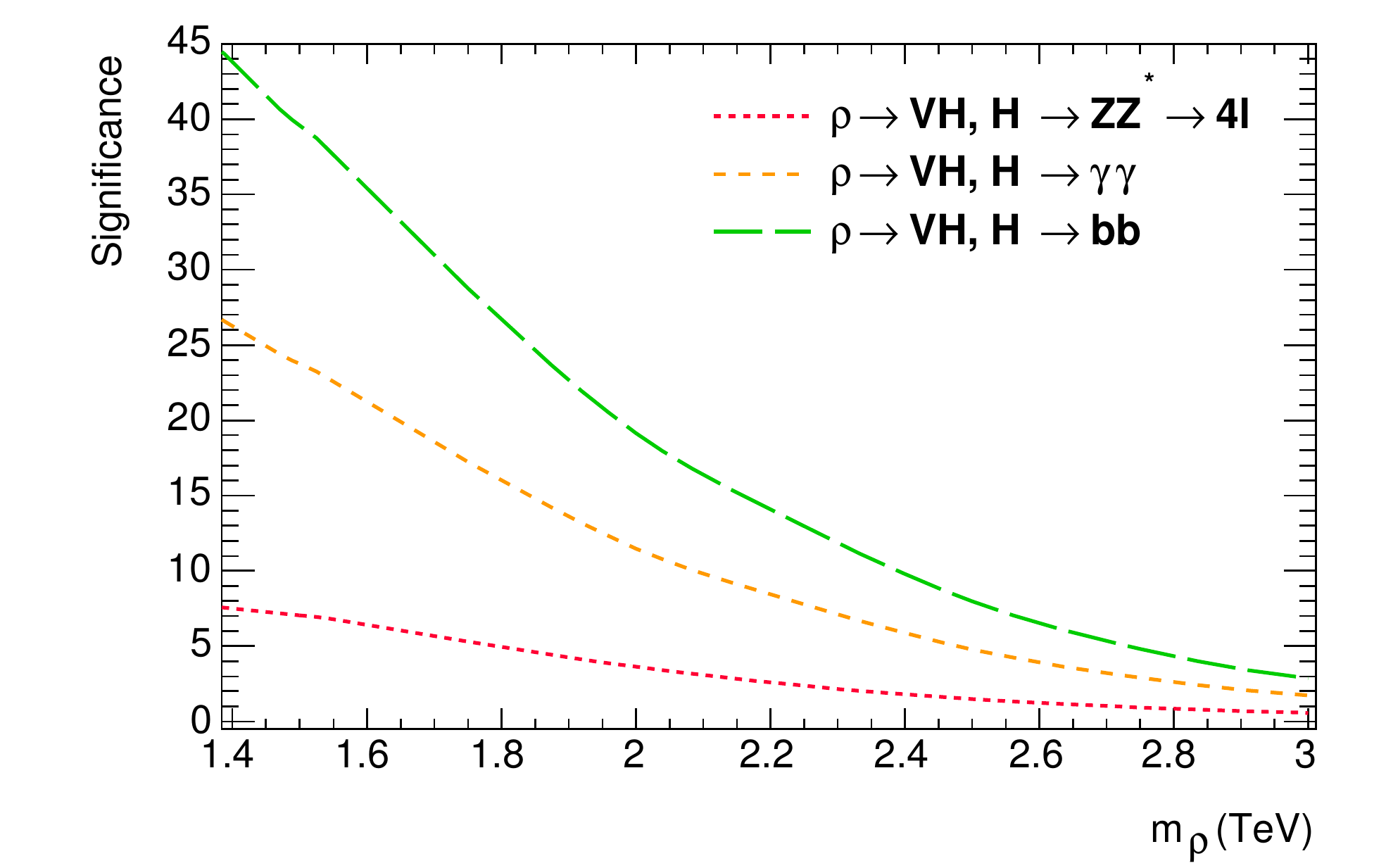}
}
\caption{
Expected significance as function of $m_{\rho}$ obtainable with a single LHC experiment using the $H\rightarrow \gamma\gamma$, $H\rightarrow ZZ^{*} \rightarrow 4\ell$ and $H \rightarrow b\bar{b}$ decay channels. The latter was obtained with an optimistic extrapolation of the results obtained with the $H\rightarrow \gamma\gamma$ channel with the results reported in \cite{ttH, Hbb}. The significance was computed with an integrated luminosity of $\mathcal{L}=3000$ fb$^{-1}$ collected at $\sqrt{s}=13$ TeV with the parameter values $\xi=\left(v/\Lambda_c\right)^2=0.1$, $g_{\rho}=4$.}
\label{signif3000}
\end{figure}

\begin{figure}[htb]
\resizebox{0.45\textwidth}{!}{
\includegraphics{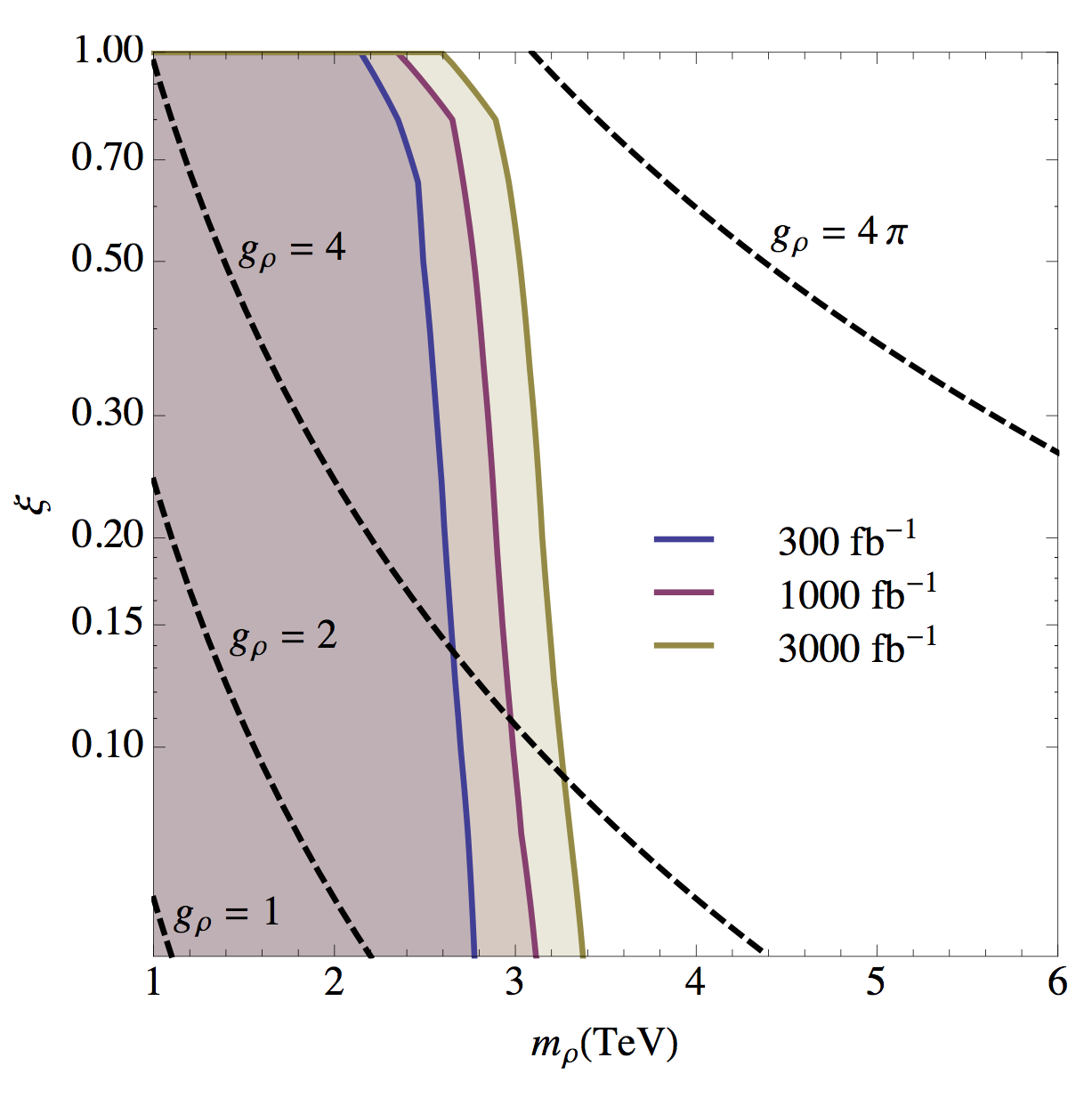}
}
\caption{
Summary of the 95\% CL expected sensitivity for a single LHC experiment at  $\sqrt{s} = 13$ TeV. The bands show the regions on the 
($\xi,m_{\rho}$) plane that can be excluded using the combination of the channels $H \rightarrow \gamma \gamma$, $H \rightarrow ZZ^{*} \rightarrow 4 \ell$ and $H \rightarrow b\bar{b}$. The latter was obtained with an optimistic extrapolation of the results obtained with the $H\rightarrow \gamma\gamma$ channel with the results reported in \cite{ttH, Hbb}. The $g_{\rho}=4\pi$ line defines the perturbativity boundary above which the effective lagrangian description breaks down.}
\label{combsen}
\end{figure}

The analysis described in Section \ref{sec:analysis} is used to assess
the compositeness discovery potential at the LHC. The results are
presented in terms of significance defined as $Z = \sqrt{2\left[ (s+b)\ln(1+s/b)- s\right]}$  \cite{asymp},
with $s$ and $b$ being the number of signal and background
events remaining after the full selection. \\
\indent The expected impact of the inclusion of the $H\rightarrow b\bar{b}$ channel is also considered.
The problem with this channel is the presence of significant QCD non-SM Higgs background
requiring careful subtraction and detailed treatment of systematics.
A realistic estimate of the impact of $H\rightarrow b\bar{b}$ can be obtained 
using existing analyses in similar event topologies to the VH.
Recent studies at the LHC \cite{ttH, Hbb} showed that when searching for a Higgs boson produced via the $t\bar{t}H$ production mechanism, 
the expected significance increases by a factor of 1.7 when combining the $H\rightarrow\gamma\gamma$ channel with $H\rightarrow b\bar{b}$. The result reported in \cite{ttH, Hbb} is based on a detailed, full simulation including all systematics. Therefore, to realistically include the impact of the $H\rightarrow b\bar{b}$ channel, the significance obtained here with the $H\rightarrow\gamma\gamma$ channel is extrapolated by the factor 1.7. \\
\indent The significance as a function of the $\rho$ mass scale ($m_{\rho}$) is considered for two integrated 
luminosity points: $\mathcal{L} = 300$~fb$^{-1}$ and $\mathcal{L} = 3000$~fb$^{-1}$. 
These sample sizes correspond to the expected integrated luminosity recorded after LHC and HL-LHC operation anticipated around
the years 2020 and 2030, respectively \cite{ECFA}. The
results obtained for the individual channels at $\sqrt{s}=13$ TeV are presented in Figure  \ref{signif300} and \ref{signif3000} for $m_{\rho}$ in the interval $1.4-3$~TeV.\\
\indent As observed in Figures \ref{signif300} and \ref{signif3000} the
diphoton channel is more sensitive than the 4-lepton. This feature is a result of the  
BR$(H\rightarrow \gamma \gamma$)  being more than an order of
magnitude larger than BR$(H\rightarrow ZZ^{*}\rightarrow 4 \ell$). The drop
in significance with increasing $m_{\rho}$ is a result
of the $\rho$ production cross section decreasing with higher
$m_{\rho}$. From Figure \ref{signif300} we see that a single LHC experiment with a data sample of 
$\mathcal{L}=300$ fb$^{-1}$ is able to observe a signal with a significance of  3$\sigma$ at 
$m_{\rho} \sim 2.5$ TeV. With  $\mathcal{L}=3000$ fb$^{-1}$ the search
is sensitive all up to $m_{\rho} \sim 3$ TeV. A combination of ATLAS and CMS measurements with $\mathcal{L}=3000$ fb$^{-1}$ 
can hence allow sensitivity to compositeness scales up to $m_{\rho} \sim 3-4$ TeV.\\
\indent The 95\% CL sensitivity expected for the combination of the three channels for the $SO(5)/SO(4)$ model considered here with $m_{\rho}\sim g_{\rho}f$ on the ($\xi$,~$m_{\rho}$) plane is presented in Figure \ref{combsen} for three different luminosity scenarios. The 
non-pertubative limit $g_{\rho}=4\pi$, where $g_\rho$ the physical coupling of the three $\rho$ resonances, is also shown.\\
\indent It is worth noting that this result might be modified if the impact of fermion resonances is taken into account. This effect is highly model-dependent and relevant only if the $\rho$ resonances interact directly with fermion resonances. In this case the interactions of vector mesons with third generation quarks can be enhanced due to partial compositeness. This does not affect the $\rho$ production cross sections, but modifies the $\rho$ decay widths into third generation quarks. Moreover, if the fermion resonances are light (which is motivated by naturalness arguments) the decays of the vector meson into a SM fermion and a fermion resonance, and the decays of $\rho$ into two fermion resonances, might become kinematically available. This would make the impact on the width and branching ratios of the vector meson even stronger. 

%
%

\section{Summary and Conclusions}
In this paper the compositeness discovery potential for the upcoming LHC runs was presented. A method was proposed to directly probe the Higgs boson compositeness 
by identifying boosted Higgs events from $\rho\rightarrow VH$ decays.
We have demonstrated that a search for compositeness is feasible
with a data sample as small as $\mathcal{L}=300$ fb$^{-1}$ collected at $\sqrt{s} = 13$ TeV, 
which corresponds to the data sample recorded by the LHC in the year 2020.\\
\indent The main result of this work is that
compositeness scales up to $g_{\rho}f\sim 3-4$ TeV can be probed at the LHC 
with $\mathcal{L} = 3000$~fb$^{-1}$ by alone using the VH decay mode. 
Combining the VH channel with dilepton and dijet searches 
could further increase the sensitivity and reach of the search.
The proposed search
can immediately be employed in the Higgs analyses currently 
performed by the ATLAS and CMS experiments.

\section*{Acknowledgements}
AK would like to thank the Mainz Institute for Theoretical Physics (MITP) for its hospitality and support during the completion of this work.

\appendix

\section{Effective Lagrangian for a $\rho$ resonance}

Following \cite{Contino:2011np}, the PNG bosons $\Pi\left( x\right) =\Pi^{\hat{a}}\left( x\right) T^{\hat{a}}$ of $\mathcal{G}\rightarrow\mathcal{H}$ symmetry breaking can be described by $U\left( \Pi\right) =e^{i\Pi\left( x\right) /f}$ transforming as
\begin{equation}
U\left( \Pi\right) \rightarrow g\; U\left( \Pi\right) \; h^{\dag}\left( \Pi , g\right) ,\ \ \ \ \ \ g\in\mathcal{G},\; h\in\mathcal{H}.
\end{equation}
The leading order effective Lagrangian term describing self-interactions of these bosons takes the form
\begin{equation}
\mathcal{L}^{\Pi }=\frac{f^2}{4}\tr\left\lbrace d_{\mu}d^{\mu}\right\rbrace 
\end{equation}
where $d_{\mu}$ is defined by
\begin{equation}
-iU^{\dag}D_{\mu}U=d^{\hat{a}}_{\mu}T^{\hat{a}}+E^{a}_{\mu}T^{a}=d^a+E^a 
\end{equation}
and $T^{\hat{a}}$, $T^a$ are the broken and unbroken generators of $\mathcal{G}$. The covariant derivative takes into account the external gauging and introduces interactions of PNG bosons with elektroweak bosons.
For the description of the vector meson transforming as
\begin{equation}
(T^{a}\rho^{a}_{\mu})\ \rightarrow\ h\; (T^{a}\rho^{a}_{\mu})\; h^{\dag}-\frac{i}{g_{\rho}}h\partial_{\mu}h^{\dag},\ \ \ \ h\in\mathcal{H} ,
\end{equation}
(where in our case $T^{a}$ are $SU(2)_L$ generators), we use the general leading-order effective Lagrangian
\begin{equation}
\mathcal{L}^{\rho}=-\frac{1}{4g_{\rho}^2}\rho^{a}_{\mu\nu}\rho^{a\;\mu\nu}+\frac{m_{\rho}^2}{2g_{\rho}^2}\left( \rho^{a}_{\mu}-E^{a}_{\mu}\right) ^2 .
\end{equation}
The connection term $E^{a}_{\mu}$ introduces interactions of $\rho$ mesons with PNG bosons and electroweak bosons.



\end{document}